\def\papertitle{Self-Supervised Disentanglement
of Harmonic and Rhythmic Features in Music Audio Signals} 
\def\paperauthorA{Yiming Wu}

\documentclass[twoside,a4paper]{article}
\usepackage{etoolbox}
\usepackage{dafx_23}
\usepackage{amsmath,amssymb,amsfonts,amsthm}
\usepackage{euscript}
\usepackage[T1]{fontenc}
\usepackage[utf8]{inputenc}
\usepackage{ifpdf}
\usepackage[english]{babel}
\usepackage{caption}
\usepackage{subfig} 
\usepackage{color}

\input glyphtounicode
\ninept

\newcounter{numauth}
\newcounter{listcnt}
\newcommand\authcnt[1]{\ifdefined#1 \stepcounter{numauth} \fi}

\newcommand\addauth[1]{
\ifdefined#1 
\stepcounter{listcnt}
\ifnum \value{listcnt}<\value{numauth}
\appto\authorslist{, #1}
\else
\appto\authorslist{~and~#1}
\fi
\fi}
\authcnt{\paperauthorB}
\authcnt{\paperauthorC}
\authcnt{\paperauthorD}
\authcnt{\paperauthorE}
\authcnt{\paperauthorF}
\authcnt{\paperauthorG}
\authcnt{\paperauthorH}
\authcnt{\paperauthorI}
\authcnt{\paperauthorJ}
\def\authorslist{\paperauthorA}
\addauth{\paperauthorB}
\addauth{\paperauthorC}
\addauth{\paperauthorD}
\addauth{\paperauthorE}
\addauth{\paperauthorF}
\addauth{\paperauthorG}
\addauth{\paperauthorH}
\addauth{\paperauthorI}
\addauth{\paperauthorJ}

\usepackage{times}

\newif\ifpdf
\ifx\pdfoutput\relax
\else
   \ifcase\pdfoutput
      \pdffalse
   \else
      \pdftrue
\fi

\ifpdf 
  \usepackage[pdftex,
    pdftitle={\papertitle},
    pdfauthor={\authorslist},
    pdfsubject={Proceedings of the 26th International Conference on Digital Audio Effects (DAFx23)},
    colorlinks=false, 
    bookmarksnumbered, 
    pdfstartview=XYZ 
  ]{hyperref}
  \usepackage[pdftex]{graphicx}
\else 
  \usepackage[dvips]{epsfig,graphicx}
  \usepackage[dvips,
    pdftitle={\papertitle},
    pdfauthor={\authorslist},
    pdfsubject={Proceedings of the 26th International Conference on Digital Audio Effects (DAFx23)},
    colorlinks=false, 
    bookmarksnumbered, 
    pdfstartview=XYZ 
  ]{hyperref}
\fi
\usepackage[hypcap=true]{caption}
\title{\papertitle}

\affiliation{
\paperauthorA\,\thanks{\vspace{-3mm}}}
{\href{https://alphatheta.com/}{AlphaTheta Corporation} \\ Yokohama, Japan\\
{\tt \href{mailto:yiming.wu@alphatheta.com}{yiming.wu@alphatheta.com}}
}

\usepackage{amsmath,amsfonts,amssymb,bm}

\newcommand{\ie}{\textit{i.e.}}
\newcommand{\eg}{\textit{e.g.}}

\newcommand{\eqdef}{\overset{\text{\fontsize{6pt}{0pt}\selectfont def}}{=}}

\newcommand{\mX}{\mathbf{X}}

\newcommand{\mZ}{\mathbf{Z}}

\newcommand{\vx}{\mathbf{x}}

\newcommand{\vz}{\mathbf{z}}

\begin{document}
\ifpdf 
  \DeclareGraphicsExtensions{.png,.jpg,.pdf}
\else  
  \DeclareGraphicsExtensions{.eps}
\fi


\maketitle

\begin{abstract}
The aim of latent variable disentanglement is to infer the
 multiple informative latent representations that lie behind
 a data generation process and is a key factor 
 in controllable data generation.
In this paper, we propose a deep neural network-based 
 self-supervised learning method to infer the
 disentangled rhythmic and harmonic representations
 behind music audio generation.
We train a variational autoencoder that generates
 an audio mel-spectrogram from two latent features 
 representing the rhythmic and harmonic content.
In the training phase,
 the variational autoencoder is trained to reconstruct the input mel-spectrogram
 given its pitch-shifted version.
At each forward computation in the training phase,
 a \textit{vector rotation} operation is applied to
 one of the latent features,
 assuming that the dimensions of the feature vectors are
 related to pitch intervals.
Therefore, in the trained variational autoencoder,
 the rotated latent feature represents
 the pitch-related information of the mel-spectrogram,
 and the unrotated latent feature represents 
 the pitch-invariant information, \ie, the rhythmic content.
The proposed method was evaluated
 using a predictor-based disentanglement metric
 on the learned features.
Furthermore, we demonstrate its application to 
 the automatic generation of music remixes.

\end{abstract}

\section{Introduction}
\label{sec:intro}

\begin{figure}[ht]
  \centerline{\includegraphics[scale=0.46]{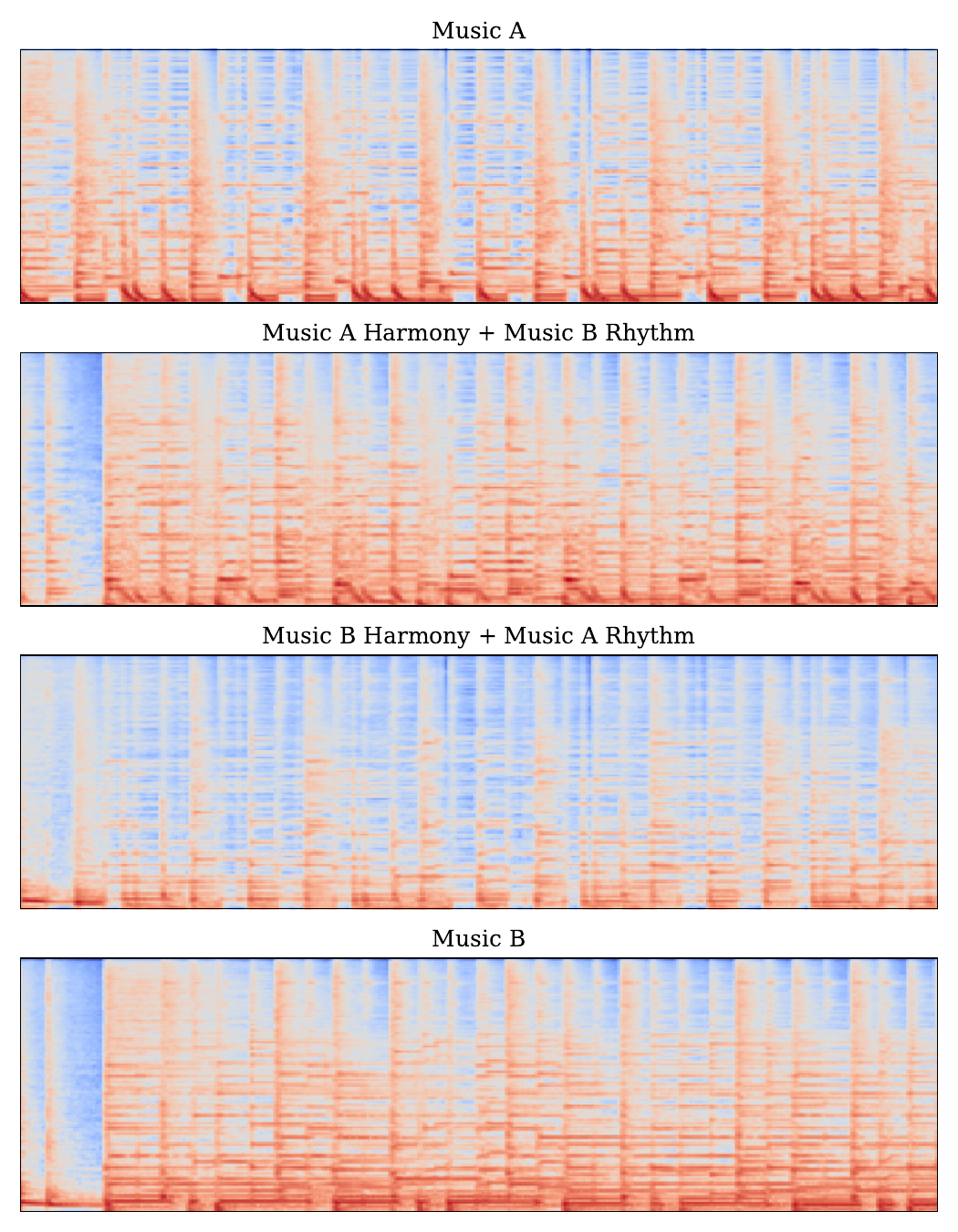}}
  \caption{\label{fft_plot}{\it An example of music remix generation by the proposed harmony-rhythm disentanglement method. The middle two spectrograms were generated by combining the harmony and rhythm contents of different music.}}
  \end{figure}

Deep neural network (DNN)-based data generation techniques
 are increasingly used in creative fields.
In the audio domain, exciting new methods have been
 proposed for speech generation, music composition,
 and sound design.
The main advantage of DNNs is their high expressiveness
 in approximating the real-world data distributions,
 which can provide consistent generation results
 that are convincing to human creators.
However, because of their highly complicated architecture,
 the interpretability and controllability of the generative process
 have become the two main problems with DNN-based data generation.
A DNN contains a huge number of stochastically optimized
 parameters,
 and hence it is impossible to explain how each
 parameter or each internal output influences the final output.
In addition, a DNN-based generative method often introduces
 a stochastic process,
 which improves the diversity of the generation results,
 but also makes it more difficult for the users
 to control the output and obtain results that reflect
 their intentions.

%


%

\textit{Disentanglement learning} is 
 a key approach to solving the problems of
 interpretability and controllability.
Disentanglement learning aims to model the generative
 process conditioned by multiple \textit{disentangled} 
 latent variables, \ie, a set of independent variables
 that are sensitive only to certain factors
 of the observed data.
For example, studies in the speech domain focus on 
 the representations of speaker identity, gender,
 speed of speech, and emotions \cite{hsu_hierarchical_2019,wang_style_2018}.
Generative models with properly 
 disentangled latent variables
 make it easier to explicitly reflect the intentions 
 of human users in the generation results.

In this paper, we focus on disentanglement learning
 for generative models of musical audio.
More specifically, our goal is to learn the disentangled
 latent features of the rhythmic and
 harmonic content in musical audio.
For human listeners, the rhythmic 
 content of a piece of music
 is derived from the onset timings of the musical audio,
 and the harmonic content is derived from the different pitches
 of the musical audio.
Therefore, these two types of content are considered to be 
 independent of each other.
In the time-frequency representations 
 (such as spectrograms) of musical audio, 
 the rhythmic and harmonic content can be
 observed in the temporal progressions along the time and
 frequency axes, respectively, and this can be used
 to implement a harmonic-percussive source
 separation algorithm \cite{fitzgerald_harmonicpercussive_2010}.
We assume that harmonic and rhythmic content
 can also be separated in latent space.

We propose a simple training method to obtain the 
 disentangled latent features
 by introducing several constraints during
 the training process.
It involves training a generative model for
 music audio spectrogram using a variational
 autoencoder (VAE),
 in which the encoder network maps the input
 spectrogram to the latent features
 while the decoder network maps the latent features
 back to the audio spectrogram.
The key idea behind our approach is to
 let the VAE not only reconstruct the input spectrogram,
 but also reverse the transformation applied to
 the input spectrogram.
In the proposed method, the transformation is 
 audio pitch-shifting.
We assume that pitch-shifting on the musical audio
 only changes its harmonic content
 and not its rhythmic content.
By introducing a vector rotation
 on the harmonic latent feature to reverse the
 pitch shift operation,
 the rotated and unrotated
 latent features can be trained without supervision
 to represent the pitch-related and pitch-invariant
 information in musical audio, respectively.

The main contribution of this work is to propose an
 effective disentanglement learning method
 that is suitable for DNN-based music audio generation models.
In the evaluation section, we show the
 quality of disentanglement
 quantitatively using a predictor-based metric.
We also explore the application of
 the proposed method to the
 automatic generation of music remixes,
 by replacing the rhythmic (or harmonic) feature
 of one musical audio clip with that of 
 another musical audio clip.
The quantitative evaluations and concrete audio examples 
 demonstrate that
 the proposed method
  can generate realistic music remixes that possesses the characteristics of both sources of music.

\section{Related Work}

This section reviews related work on DNN-based 
 generation and disentanglement learning for
 musical audio.

\subsection{DNN-based Musical Audio Generation}

Several different approaches have been proposed for
 DNN-based music audio generation.
One popular approach is based on differential
 digital signal processing (DDSP) \cite{engel_ddsp_2020},
 in which the generative model is concatenated with
 audio DSP modules such as filters and oscillators.
DNNs are then trained to estimate the 
 parameters of these DSP modules.
Because DDSP-based generative models 
 utilize strong inductive biases,
 they are generally more interpretable,
 and require fewer audio examples to achieve reasonable
 generalized performance.
Therefore, DDSP has been applied in several existing
 synthesizer algorithms,
 such as wavetable synthesizer \cite{shan_differentiable_2022},
 waveshaping synthesizer \cite{hayes_neural_2021}, 
 FM synthesizer \cite{caspe_ddx7_2022}, 
 and the WORLD vocoder \cite{nercessian_differentiable_2022}. 
 
Another approach is the autoencoding approach,
 which trains a DNN-based generative
 model and its latent feature space
 using an autoencoder network.
Once the autoencoder has been trained,
 musical audio can be generated by manipulating the
 latent feature and reconstructing the audio
 using the decoder network.
More specifically, one can interpolate over the latent feature
 space like RAVE \cite{caillon_rave_2021},
 or train the language model of the latent feature
 to generate musical audio from scratch,
 as in Jukebox \cite{dhariwal_jukebox_2020}, 
 Musika \cite{pasini_musika_2022}, and MusicLM \cite{agostinelli_musiclm_2023}.

\subsection{Disentanglement Learning for Audio}

The main goal of disentanglement learning for audio
 is to implement audio transformation systems that
 change certain aspects of the musical content,
 such as timbre or musical styles.
For example, Noam et al. proposed a
 music translation method that transforms
 the domain (musical instruments and styles) of
 musical audio~\cite{mor_universal_2019}.
The method is based on a multi-domain autoencoder
 based on WaveNet \cite{oord_wavenet_2016}, 
 where the encoder WaveNet transforms
 the audio waveform into a domain-independent latent
 representation, and the domain-specific WaveNet
 decoders reconstruct the audio waveform from
 the latent representation.
To make the encoder extract the domain-independent
 representation from audio waveforms,
 the encoder is trained to fool a domain classifier network
 that tries to correctly recognize the domain type
 from the latent representation.
This approach is not a fully unsupervised method because
 a domain label should be given for each musical audio clip
used to train the neural networks.

Studies on disentanglement learning for audio
 have proposed several learning schemes to automatically
 separate the pitch-related and pitch-invariant
 information in the musical audio in the latent space of an audio generative model.
Luo et al. proposed 
 a learning method to encode the pitch and timbre of
 musical instrument sounds using Gaussian mixture VAE \cite{luo_learning_2019} ,
 where the latent representations were learnt 
 in a supervised and semi-supervised manner
 using pitch and instrument annotations.
GANStrument proposed by Narita et al. 
 introduces an adversarial training scheme
 to extract pitch-invariant features from musical
 instrument sound \cite{narita_ganstrument_2023}.
Using the trained feature encoder,
 GANStrument can generate pitched instrument
 sounds given a one-shot sound as input.
Luo et al. also 
 proposed an unsupervised learning method
 to encode the pitch and timbre of
 musical instrument sounds,
 in which the pitch is represented as a discrete label
 and the timbre is represented as a continuous
 feature vector \cite{luo_unsupervised_2020} .
Similar to our proposed method, they assume
 that a moderate pitch shift operation
 does not change the timbre of the original
 musical instrument sound.
Based on this assumption,
 they treat the original sound and its pitch-shifted
 version as a pair,
 and swap the encoded pitch variables before 
 reconstructing the musical sound using the decoder.
Because the pitch is represented as a single
 discrete variable, this method is suitable for
 monophonic musical sound.
Our proposed method formulates a VAE in a similar way;
 however, we formulate the pitch-related feature 
 as continuous value vectors, so that these vectors
 can represent the polyphonic pitch information
 found in any kind of musical recording.

 \begin{figure*}[ht]
  \center
  \includegraphics[width=6.7in]{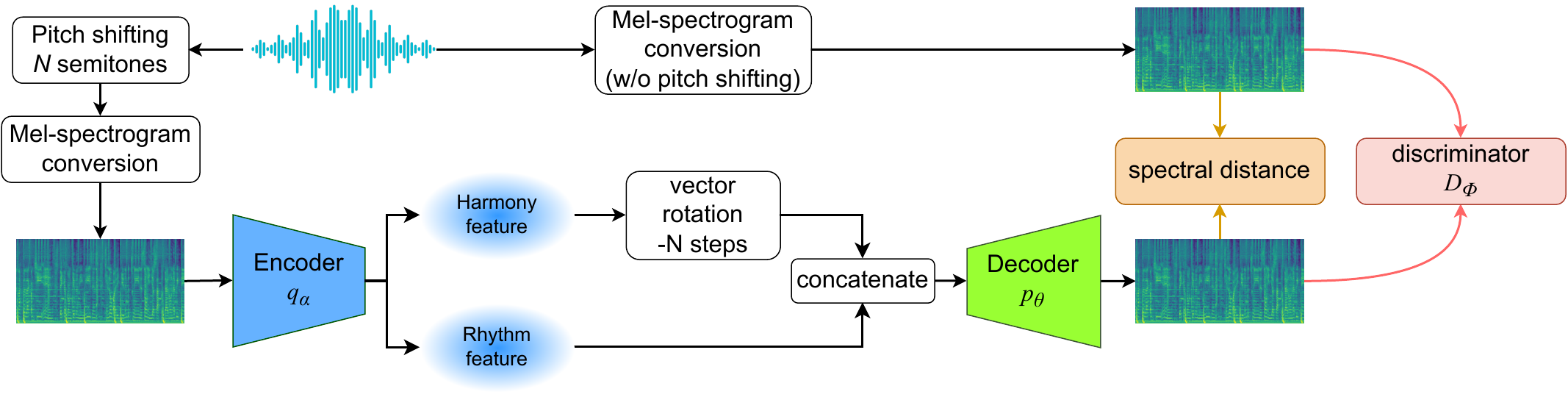}
  \caption{\label{fig:overall_architecture}{\it Proposed VAE architecture and its forward computation procedure.}}
  \end{figure*}

\section{Proposed Method}

This section describes the proposed
 self-supervised disentanglement learning method.
An overview of the proposed method
 is shown in Fig. \ref{fig:overall_architecture}.
We formulate a probabilistic generative model
 representing the generative process of
 an audio mel-spectrogram from two latent features
 representing harmony and rhythm
 in the form of a VAE
 (Section \ref{subsect:vae}).
In the training phase, we use an audio
 pitch-shifting algorithm to enable the model
 to learn the two
 latent features that represent the pitch-related
 and pitch-invariant information of the input audio
 (Section \ref{subsect:learning}).
In addition, we train the decoder as a generative
 adversarial network (GAN) to improve the generation quality
 (Section \ref{subsect:adversarial}).

\subsection{VAE Formulation}
\label{subsect:vae}

Let $\mX=\{\vx_n\}_{n=1}^N$ be a log-scaled mel-spectrogram of a musical audio,
 represented as a sequence of D-bins spectrum $\vx_n\in \mathcal{R}^D$.
Let $\mZ^h=\{\vz^h_n\}_{n=1}^N$ and $\mZ^r=\{\vz^r_n\}_{n=1}^N$ be sequences of latent features,
 where $\vz^h_n, \vz^r_n\in \mathcal{R}^L$ are $L$-dimensional continuous-valued vectors ($L=128$)
 that abstractly represent the harmonic and rhythmic
 content at the $n$th audio frame, respectively.
We formulate a generative model with $\mX$ as the
 observed variable and $\mZ^h$, $\mZ^r$ as
 the latent features as follows:
\begin{equation}
  p(\mX) = p_\theta(\mX|\mZ^h, \mZ^r)p(\mZ^h)p(\mZ^r)
  \label{eq:generative_model}
\end{equation}
where $p_\theta(\mX|\mZ^h, \mZ^r)$ is a conditional generative model with parameters $\theta$.
We define $p_\theta$ as a decoder neural network
 parametrized by $\theta$.
The decoder network models the generative process of mel-spectrogram
 from the two latent features $\mZ^h$ and $\mZ^r$. 
 In our work, we evaluate $p_\theta(\mX|\mZ^h, \mZ^r)$ using
 the spectral distance between $\mX$ and
 the output of the decoder network $\omega_\theta(\mZ^h, \mZ^r)$:
 \begin{equation}
  p_\theta(\mX|\mZ^h, \mZ^r)\sim S_\theta(\mX,\mZ^h, \mZ^r)\eqdef ||\mX-\omega_\theta(\mZ^h, \mZ^r)||_1
  \label{eq:specdistance}
 \end{equation}
where $||\cdot||_1$ is the \textit{L1} norm.

Since the inference model of the latent features $p(\mZ^h, \mZ^r|\mX)$ is intractable,
 we use a neural encoder network $q_\alpha$
that approximates the distributions 
 of the latent features given an observed mel-spectrogram
 as follows:
\begin{equation}
  q_\alpha(\mZ^h | \mX) = \prod_{n=1}\mathcal{N}(\vz^h_n|\mu_\alpha(\mX)_n^h,\sigma_\alpha(\mX)_n^h)
  \label{eq:encoder_rhythm}
\end{equation}
\begin{equation}
  q_\alpha(\mZ^r | \mX) = \prod_{n=1}\mathcal{N}(\vz^r_n|\mu_\alpha(\mX)_n^r,\sigma_\alpha(\mX)_n^r)
  \label{eq:encoder_harmony}
\end{equation}
where $\mu_\alpha(\mX)^h$, $\sigma_\alpha(\mX)^h$,$\mu_\alpha(\mX)^r$, and $\sigma_\alpha(\mX)^r$ are
 the four parts of the encoder network output.

The priors $p(\mZ^h)$ and $p(\mZ^r)$ are
 set to a standard Gaussian distribution as follows:
 \begin{align}
  p(\mZ^r)
  =
  \prod_{n=1}^N \mathcal{N}(\vz^h_n | \mathbf{0}_L, \mathbf{I}_L), \\
  p(\mZ^h)
  =
  \prod_{n=1}^N \mathcal{N}(\vz^r_n | \mathbf{0}_L, \mathbf{I}_L),
  \label{eq:p_z}
 \end{align}

\begin{figure*}[ht]
  \center
  \includegraphics[width=6.5in]{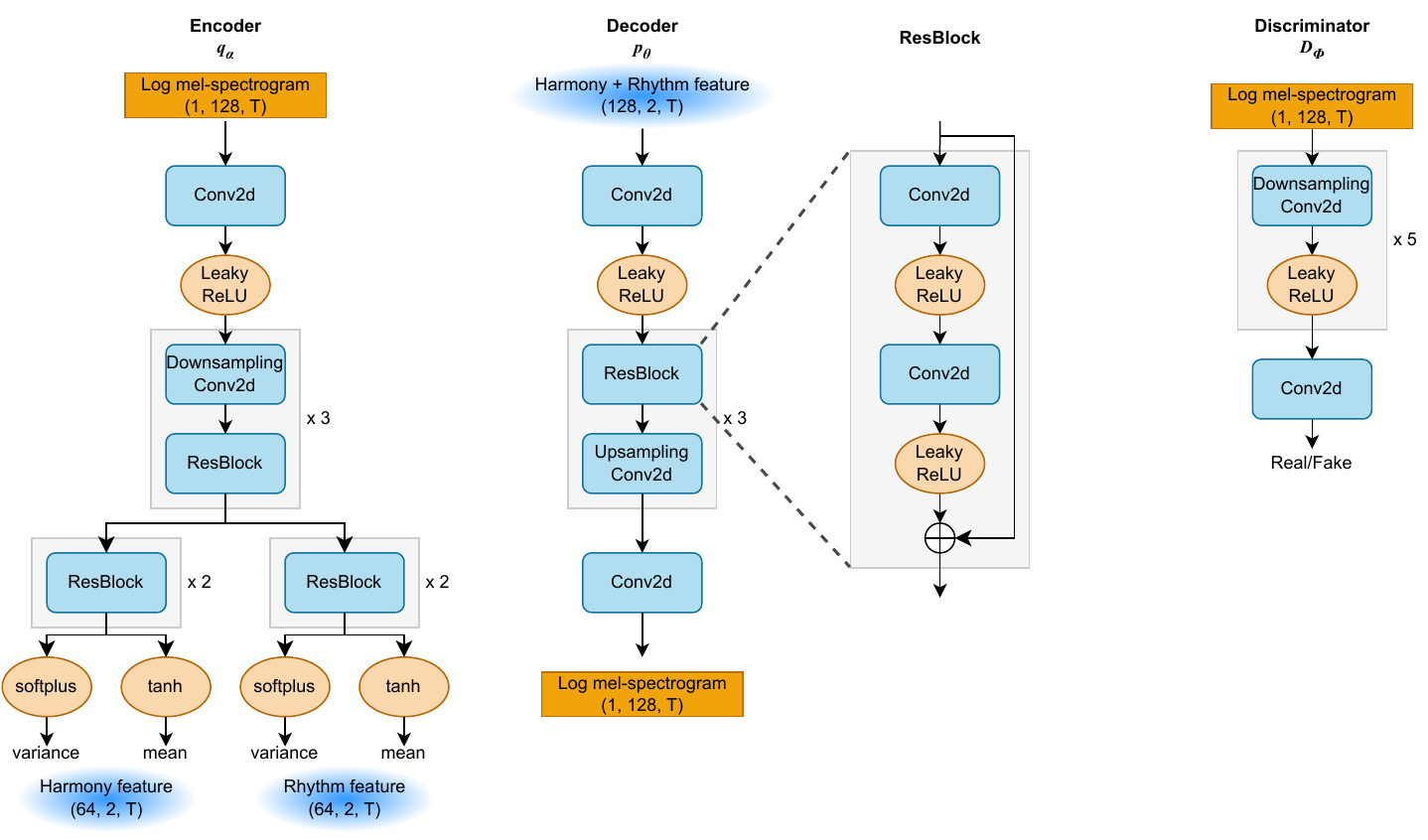}
  \caption{\label{fig:network_specification}{\it Proposed VAE architecture and its forward computation procedure.}}
  \end{figure*}

As shown in Fig.\ref{fig:network_specification},
 the encoder neural network is composed of 
 stacked residual convolution layers and
 downsampling layers.
Two independent bottleneck modules are
 appended to the bottom layer of the encoder
 to compute the parameters of the two 
 latent distributions.
Each downsampling layer is implemented
 with a strided convolution
 layer that reduces the dimension of the frequency axis
 of the mel-spectrogram by a factor of four
 while keeping the dimension of the time axis unchanged.
Therefore, the encoder reduces the frequency axis of the input
 spectrogram by a factor of 64, and outputs the latent features
 with two dimensions on the frequency axis.
Similarly, the decoder neural network is composed
 of stacked residual convolution layers and
 upsampling layers that are implemented with
 strided transposed convolution layers,
 each of which expands the frequency-axis by a factor
 of four.

\subsection{Self-Supervised Disentanglement Learning}
\label{subsect:learning}

In a normal VAE setting \cite{kingma_auto-encoding_2014}, the generative model is trained within the framework of
 variational inference,
 which jointly optimizes the encoder 
 and decoder network
 to maximize the evidence lower bound (ELBO) 
 of the observed data
 likelihood $p(x)$ as:
\begin{align}
  &\mathcal{L}_{VAE_{normal}}=
  \mathbb{E}_{q_\alpha(\mZ^r,\mZ^h|\mX)}[\log p_\theta(\mX|\mZ^r, \mZ^h)]\nonumber\\
  &-\beta\mathcal{D}_{KL}(q_\alpha(\mZ^h|\mX)||p(\mZ^h))-\beta\mathcal{D}_{KL}(q_\alpha(\mZ'^r|\mX)||p(\mZ^r))
  \label{eq:ord_vae_objective}
\end{align}
where $\mathcal{D}_{KL}(q||p)$ is the KL divergence from distribution $q$ to $p$, 
 and $\beta$ is a weighting factor that controls the trade-off
between the reconstruction accuracy and
level of disentanglement within the latent features \cite{higgins_beta-vae_2017}.
The latent variable regularization term in ELBO encourages
 disentanglement between each dimension of the latent variable\cite{higgins_beta-vae_2017}.
However, without explicit conditioning,
 there is no guarantee that the latent variables learn to explicitly represent
 the harmonic (or rhythmic) aspects of the mel-spectrogram.

To distinguish the harmonic and rhythmic content of 
 musical audio, we make the assumption that
 rhythmic content is invariant to audio pitch shifting,
 whereas harmonic content is not.
Assuming that the musical audio share the same
 tuning (\eg, tuned to 440Hz),
 we add a definition of the dimensions of the
 latent vector $z_n^h$: the $i$-th dimension $z_{n_i}^h$
 represents the pitch information of a certain pitch
 height, and the pitch intervals between
 the pitches corresponding to the $i$-th and $j$-th
 dimension is $j-i$ times of a small pitch interval unit
 (we use \textit{semitone} in the following statements).
In this way, we can relate audio pitch-shifting to 
 a \textit{vector rotation} operation on $z_n^h$,
 \ie, when $\mZ^h$ is the harmony feature
 of $\mX$, the $n$-step vector rotation of $\mZ^h$
 is the harmony feature of the
 $n$-semitone pitch shift version of $\mX$.

Based on our definition of $\mZ^h$, we designed a
 training procedure to facilitate the harmony-rhythm
 disentanglement.
Concretely, each forward computation in a 
 training iteration proceeds follows:
\begin{enumerate}
  \item Shift the pitch of the input audio segment by a random number of semitones $n\in [-8,8]$. Let $\mX'$ be the mel-spectrogram of the pitch-shifted audio,
  \item Calculate the latent feature distribution $q_\alpha(\mZ'^h, \mZ'^r|\mX')$ using the encoder network,
  \item Sample the latent features $\mZ'^h, \mZ'^r$ from $q_\alpha(\mZ'^h, \mZ'^r|\mX')$ using the reparameterization trick \cite{kingma_auto-encoding_2014},
  \item Apply $(-n)$-step vector rotation to the channel dimension of $\mZ'^h$. Let $\mZ^h$ be the rotated latent feature.
  \item Reconstruct the mel-spectrogram from $p_\theta(\mX|\mZ'^h, \mZ^r)$ using the decoder network.
\end{enumerate}
 
Combining Equation \ref{eq:ord_vae_objective} with Equation \ref{eq:specdistance}, the training objective of the VAE is:
\begin{align}
  &\mathcal{L}_{VAE}=
  \mathbb{E}_{q_\alpha(\mZ'^h,\mZ^r|\mX')}[S_\theta(\mX,\mZ'^h,\mZ^r)]\nonumber\\
  &-\beta \mathcal{D}_{KL}(q_\alpha(\mZ'^h|\mX')||p(\mZ^h))-\beta \mathcal{D}_{KL}(q_\alpha(\mZ'^r|\mX)||p(\mZ^r))
  \label{eq:vae_objective}
\end{align}
We set $\beta=0.1$ in our experiment, which places more weight on 
 the reconstruction accuracy.
The expectation term is approximated by the
 Monte Carlo method 
 using the reparameterization trick.
Intuitively, the VAE decoder is 
 trained to reconstruct the
 original mel-spectrogram $\mX$ given the latent variables
 encoded from the pitch-shifted mel-spectrogram $\mX'$.
Because $\mZ'^r$ is not altered during the 
 forward computation, it should represent
 the pitch-invariant elements in $\mX'$ and $\mX$.
By contrast, because the vector rotation
 on $\mZ'^h$ reverts the pitch shift on $\mX$,
 the rotated variable $\mZ^h$ is able to represent the 
 pitch-specific elements of the original $\mX$.
Therefore, unlike $\mZ^r$, $\mZ^h$ should
 represent the pitch-related elements in $\mX$
 during the optimization.

\subsection{GAN Learning}
\label{subsect:adversarial}

To improve the quality of the
 generated mel-spectrogram,
 the VAE networks are also trained as a GAN \cite{goodfellow_gan_2014}.
We additionally define a discriminator network $D_\phi$ that 
 learns to distinguish the generated mel-spectrogram
 from the original mel-spectrogram.
The GAN training objective is defined as follows:
\begin{equation}
  \mathcal{L}_{dis} = (1-D_\phi(\mX))^2 + D_\phi(\hat{\mX})^2
\end{equation}
\begin{equation}
  \mathcal{L}_{gen} = -D_\phi(\hat{\mX})^2
\end{equation}
where $\mX$ is the original spectrogram and $\hat{\mX}$
 is the spectrogram reconstructed by the VAE.
To stabilize the adversarial training process, a feature 
 matching loss $\mathcal{L}_{FM}$ \cite{kumar_melgan_2017} is further added
 to the training objective.
Altogether, the objective function for the VAE network
 optimization is
\begin{equation}
  \mathcal{L}_{total} = \mathcal{L}_{VAE}
  \nonumber
   + \mathcal{L}_{gen}+\mathcal{L}_{FM}
 \end{equation}
Following the ordinary GAN training procedure,
 the discriminator network is trained to minimize
 $\mathcal{L}_{dis}$, and the VAE network is trained to
 minimize $\mathcal{L}_{total}$.
As illustrated in Fig. \ref{fig:network_specification},
 the discriminator network is composed of five
 convolutional layers with leaky ReLU activation.

Combination of VAE and GAN objectives is 
 also used to train the RAVE \cite{caillon_rave_2021} and Musika! \cite{pasini_musika_2022} audio synthesizer.
Unlike RAVE, our method does not optimize the 
 VAE and GAN objectives separately. 
We also do not fix the parameters of the 
 encoder network. 
In our experiments, the objective 
 $\mathcal{L}_{total}$ jointly
 optimizes the encoder and decoder network.

\section{Evaluation}

This section reports the comparative experiment
 conducted to evaluate the effectiveness
 of the proposed disentanglement learning method.
The experiments were implemented using PyTorch \cite{paszke_automatic_2017},
 and the source code is available on GitHub. \footnote{\url{https://github.com/WuYiming6526/HARD-DAFx2023}}

\subsection{Datasets}

We use the \textit{fma-large} subset of the
 Free Music Archive (FMA) dataset \cite{defferrard_fma_2017} to
 train the VAE.
The dataset contains 30-second musical audio snippets 
 from 106,574 Creative Commons-licensed music tracks.
To measure the quality of the
 rhythm--harmony disentanglement of the proposed method,
 we use the RWC-Popular dataset \cite{goto_rwc_2002} as the
 test set.
The RWC-Popular dataset contains
 100 pieces of popular song audio with chord progression
 annotations.
Following the common automatic chord estimation setting,
 the annotated chord labels are reduced to the
 \textit{major} and \textit{minor} triads.

The mel-spectrogram was computed from  the audio signal
 using a sample rate of 22,050Hz.
The FFT size, window length, and hop size of the
 short-time Fourier transform 
 were set to 2048, 2048, and 512 samples, respectively,
 and the number of mel frequency bins was set to 128 (thus $D=128$).
Hann window was used for FFT computation.

A general-purpose audio pitch-shifting algorithm was used
 to obtain the pitch-shifted versions 
 of the musical audio.
In our experiments, we used the pitch-shifting function
 implemented in the \textit{Pedalboard} audio processing library, \footnote{\url{https://spotify.github.io/pedalboard/reference/pedalboard.html}}
 which wraps the \textit{Rubber Band} audio stretching library. \footnote{\url{https://breakfastquay.com/rubberband/}}
The \textit{Rubber Band} audio stretching algorithm
 is based on the phase-vocoder method 
 that uses phase resets on the percussive transients,
 an adaptive stretch ratio between phase reset points, 
 and a "lamination" method to improve vertical phase coherence.
In contrast to the naive phase-vocoder time stretching algorithm
 implemented in \textit{librosa} \cite{mcfee_librosa_2015} and \textit{torchaudio},
 \textit{Rubber Band}'s algorithm can preserve 
 percussive sounds without noticeable distortion.

\subsection{Evaluation Metrics}

We use a predictor-based evaluation 
 metric similar to that used in 
 \cite{luo_unsupervised_2020} to
 measure the disentanglement
 between the inferred rhythm and harmony features.
Specifically, 
 a sequence classification model based on
 a two-layer bidirectional gated recurrent unit (GRU) network was trained
 to predict the chord labels and onset states from the audio features $\mZ^h$, $\mZ^r$, or the original
 audio mel-spectrogram $\mX$.
The accuracy of chord label prediction was measured by the
 frame-wise label overlap rate,
 and the accuracy of onset prediction was measured by the
 binary F-1 score over the onset positions.
 
The accuracy of chord prediction and onset prediction
 measures how well the latent features
 reflect the pitch-related and pitch-invariant 
 information of the audio, respectively.
If $\mZ^r$ and $\mZ^h$ are well disentangled,
 the classifiers on
 $\mZ^r$ should yield high accuracy for onset prediction
 and low accuracy for chord label prediction.
Similarly, the classifiers on $\mZ^h$ should yield
 high accuracy for chord label prediction
 and low accuracy for onset prediction.

The RWC-Popular dataset is divided into a training set (90\%)
 and an evaluation set(10\%).
The data pairs of musical audio and chord label annotations
 in the RWC-Popular dataset were used to train and
 evaluate the chord label classifier.
Similarly, the musical audio and onset label data pairs
 were used to train and evaluate the onset label
 classifier,
 where the onset label was inferred from the raw music
 audio using the onset detection algorithm implemented
 in the \textit{librosa} library.

We further explore the application of the proposed method
 to the automatic generation of music remixes.
To generate music remixes,
 we used the trained VAE to
 generate audio spectrograms that simultaneously contain
 the musical elements of two different music tracks.
Given two pieces of beat-synchronized music A and B,
 a remix was created by the following process:

\begin{enumerate}
  \item Infer the latent representations $\mZ^h_A, \mZ^r_A, \mZ^h_B$, and $\mZ^r_B$ of the mel-spectrograms $\mX_A$ and $\mX_B$ using the encoder network,
  \item Generate the mel-spectrogram from $\mZ^h_A, \mZ^r_B$ using the decoder network.
\end{enumerate}

We used the Fr\'echet Inception Distance (FID) \cite{heusel_gans_2017} to 
 quantitatively measure the quality of the generated
 spectrograms.
The FID measure is given by:
\begin{equation}
  F(\mathcal{N}_b,\mathcal{N}_e)=||\mu_b-\mu_e||^2+tr(\Sigma_b+\Sigma_e-2\sqrt{\Sigma_b \Sigma_e})
\end{equation}
where $\mathcal{N}_b(\mu_b,\,\Sigma_b)$ is the multivariate
 normal distribution estimated from the Inception V3 \cite{szegedy_rethinking_2016}
 features calculated from a set of 
 spectrograms of the real musical audio,
 and $\mathcal{N}_e(\mu_e,\,\Sigma_e)$ is the distribution
 calculated from the generated spectrograms.
The generated spectrograms are considered to be
 more musically realistic if the computed FID is low.
The feature extractor is a pre-trained music genre
 classifier that wass trained
 using the genre-annotated musical audio
 in the FMA dataset.

We used the following music remixing methods as the baselines:
\begin{itemize}
  \item \textbf{HPSS}. We apply the harmonic-percussive source separation (HPSS) algorithm \cite{driedger_extending_2014} in the \textit{librosa} library to music A and music B, and mix the harmonic part of A and percussive part of B to create the remix version. The HPSS algorithm infers the spectral masks for harmonic and percussive parts using median-filtering along the time and frequency axis.
  \item \textbf{ASAP}. We use the \textit{Spectral Morphing} audio effect implemented in the \textit{ASAP} plug-in suite developed by IRCAM. \footnote{\url{https://forum.ircam.fr/projects/detail/asap/}} The \textit{Spectral Morphing} plugin combines the spectral characteristics of two audio signals using the source-filter technique where the audio signal of music B is used as a filter of the audio signal of music A. More specifically, the frequency-domain amplitude of the two audio signals are multiplied, while preserving the phase of the source audio. The spectral envelope of music B is further applied to the filtered signal. We set music A as the main input, music B as the sidechain input, and set the \textit{Global Mix} parameter to 100\% to generate the remixed version.
\end{itemize}

We randomly chose 20 songs from the
 RWC-Popular dataset to create 10 pairs of 
 audio clips.
Each audio clip was time-stretched to 120 BPM
 and was 8s long.
Therefore, the FID for each compared method
 was computed on 10 audio clips generated by
 the corresponding method.
The remixes created by the 
 proposed and the baseline methods
can be found on the online 
project page. \footnote{\url{https://wuyiming6526.github.io/HARD-demo/}}
Hifi-GAN \cite{kong_hifi-gan_2020} was used to convert the
 mel-spectrograms generated by the proposed 
 method into an audio signal.

\subsection{Results}

\begin{table}[]
  \centering
  \caption{\it Harmony-rhythm disentanglement Metrics}
  \label{tab: disentangle_results}
  \begin{tabular}{|c|c|c|}
  \hline
  \multicolumn{1}{|c|}{\textbf{Feature}} &
    \multicolumn{1}{c|}{\textbf{\begin{tabular}[c]{@{}c@{}}chord\\ accuracy\end{tabular}}} &
    \multicolumn{1}{c|}{\textbf{\begin{tabular}[c]{@{}c@{}}onset\\ F1\end{tabular}}} \\ \hline
  harmony feature & \textbf{69.61}\% & 60.09\% \\
  rhythm feature  & 24.65\% & \textbf{66.04}\%  \\
  mel-spectrogram & 51.95\% & 65.19\% \\ \hline
  \end{tabular}
  \end{table}

  \begin{table}[]
    \centering
    \caption{\it FIDs of the generated spectrogram}
    \label{tab:fid}
    \begin{tabular}{|c|c|}
    \hline
    \textbf{Model} & \textbf{FID}   \\ \hline
    HPSS           & 12.84          \\
    ASAP           & 13.18          \\
    Disentangled VAE (proposed)       & \textbf{12.46} \\ \hline
    \end{tabular}
    \end{table}

Table \ref{tab: disentangle_results} compares the accuracy of 
 chord classification and onset detection
 for different audio features.
The overall chord classification accuracy for the
 harmony feature was much higher than for the rhythm feature.
The chord labels were almost unpredictable from the rhythm features
 because these features were trained to be pitch-invariant.
By contrast, the beat detection score 
 was higher for the rhythm features
 than for the harmony features by a much smaller margin.
Although the rhythm features were better at 
 representing onset information,
 the harmony features were not completely onset-invariant.
This is somewhat inevitable, since onsets can be  
 inferred in part from pitch transitions.
Interestingly, both harmony and rhythm features
 scored higher than the mel-spectrogram
 representation in the chord classification 
 and onset detection tasks, respectively.
Since the latent features enhance the pitch-related and pitch-invariant
 elements in the musical audio, 
 it is reasonable that the latent features were 
 found to be more suitable for
 the pitch-related or rhythm-related music information retrieval tasks.
This result indicates that the proposed method can
 also be used as a self-supervised pre-training method
 to provide better feature representations for 
 other music information retrieval tasks.

\begin{figure}[]
  \centerline{\includegraphics[scale=0.33]{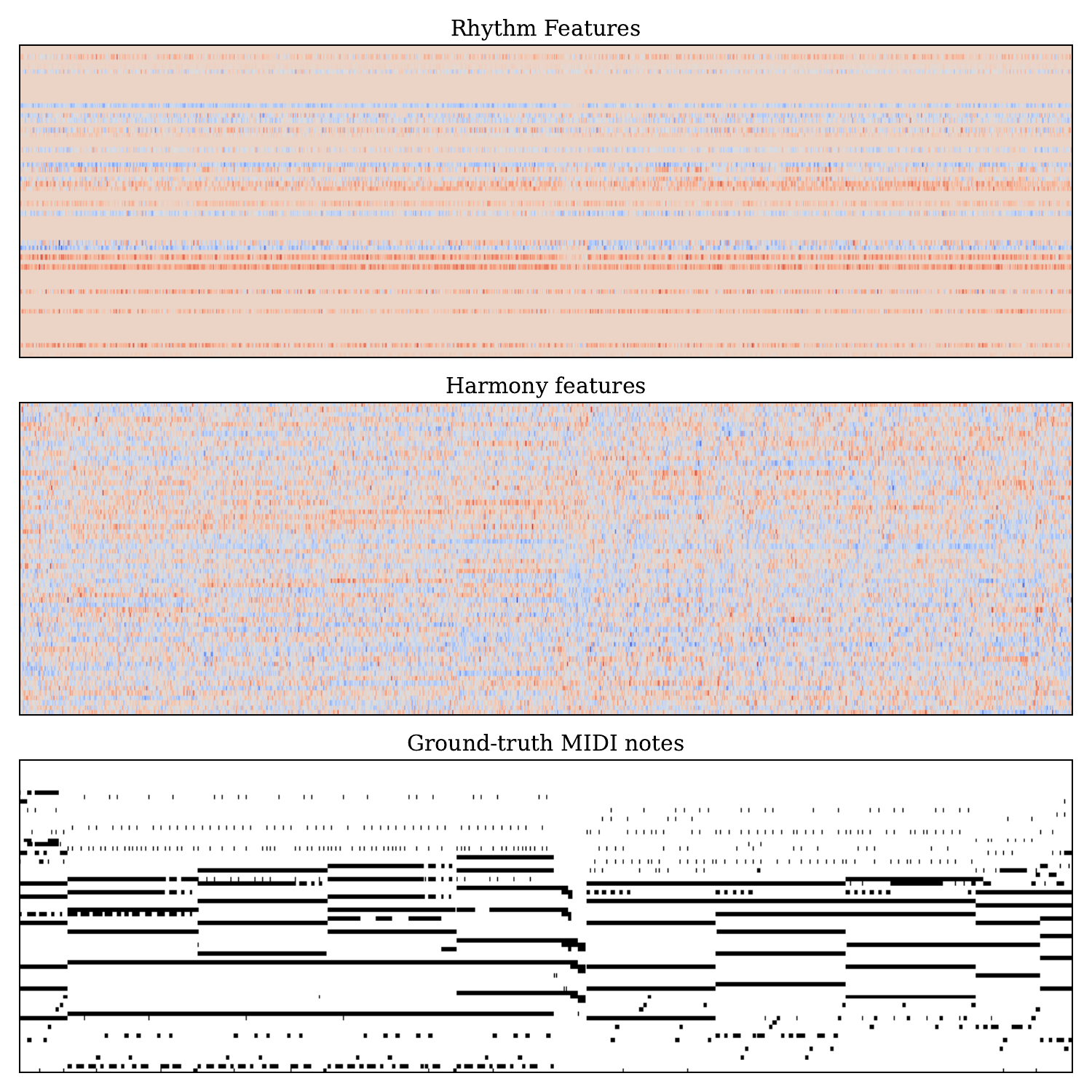}}
  \caption{\label{fig:latent_visualization}{\it Visualizations of the rhythm and harmony features of a song from the RWC-Popular dataset. The bottom figure visualizes the MIDI notes from the ground-truth MIDI file.}}
  \end{figure}

As a qualitative evaluation, we visualized
 the latent harmony and rhythm 
 representations.
Fig. \ref{fig:latent_visualization} compares the
 visualized latent features of a song from the
 RWC-Popular dataset with the ground-truth MIDI pianorolls.
It can be seen that
 the harmony feature had similar pitch progressions to
 the ground-truth pianoroll.
The rhythm features were relatively sparse,
 and there was no obvious correlation with
 the pitches or onsets of the ground-truth pianoroll.

As shown in Table \ref{tab:fid}, the remix
 generated by the proposed method achieved a better
 FID score than the baseline methods,
 suggesting that 
 the proposed method generated spectrograms
 that are closer to real audio spectrograms
 than the baseline methods.
This is a promising result as it indicates
 that the proposed method has the potential
 to generate high-quality remixes.
The \textbf{HPSS} method simply replaced the percussive
 part of music A with music B,
 so the harmonic part does not change.
The \textbf{ASAP} method added some rhythmic
 elements of music B to the audio of music A
 through the dynamic filtering effect,
 but the rhythmic sounds of music A were still 
 present.
In contrast to these baseline methods,
 the proposed method reflected the rhythmic
 elements of music B more clearly.
Unlike the results of the \textbf{HPSS} method, 
 all of the generated audio,
 including the harmonic part of music A,
 reflect the rhythm of music B.
Unlike the results generated by the \textbf{ASAP} method,
 the rhythm of music A was removed
 and only the rhythm of music B was present.

\section{Conclusion}

We proposed a simple self-supervised learning method 
 for inferring the disentangled rhythm and 
 harmony features of musical audio.
Through quantitative metrics and qualitative observations,
 we showed that the rhythm and harmony features
 obtained using the proposed method
 achieved a high degree of disentanglement.
We also demonstrated its potential use for the automatic generation of music remixes.

The generative models that can be
 used in the proposed method 
 are not limited to spectrogram-based models.
In principle, the disentanglement learning strategy
 can be applied to any kind of autoencoder-based
 audio generation model,
 including time domain-based generative models
 such as RAVE and SoundStream \cite{zeghidour_soundstream_2022}.
However, the relationship between the time-domain audio 
 signal and the audio pitch shift is less clear
 than it is in the time-frequency audio representations.
Therefore, disentanglement learning using time domain audio
 signals may be practically more challenging.
In our initial experiments, disentanglement learning
 on the time-domain generation models
 did not perform as well as it did with
 the mel-frequency domain model. 
The solution to this problem is left for 
 future research.

We also believe that the application 
 of the proposed generative model is not limited to
 music audio generation.
The proposed method could potentially
 be a pre-training method for
 downstream music information retrieval tasks.
For example,
 the disentangled acoustic representation
 of harmony and rhythm may be suitable
 for musical notes, chords, or beat transcription tasks.
Combining the encoder of the proposed VAE
 with the music transcription model
 would be worth exploring to push the boundaries of the
 automatic music transcription research.
\section{Acknowledgment}

This work has been supported by AlphaTheta Corporation.
We thank Kimberly Moravec, PhD, from Edanz (https://jp.edanz.com/ac) for editing a draft of this manuscript . 

\nocite{*}
\bibliographystyle{IEEEbib}
\bibliography{refs} 

\end{document}